\documentclass[prl,amsmath,amssymb,twocolumn,superscriptaddress]{revtex4-2}
\usepackage{amsmath}
\usepackage{amssymb}
\usepackage{amstext}
\usepackage{amsfonts}
\usepackage{amsxtra}
\usepackage{bm}
\usepackage[usenames]{color}
\usepackage{grffile}
\usepackage{soul}
\usepackage{epstopdf}
\usepackage{xcolor}
\usepackage{graphicx}
\usepackage{marvosym}
\usepackage{wasysym}
\usepackage{physics}

\usepackage[caption=false]{subfig}

\usepackage[colorlinks=true, letterpaper=true, pdfstartview=FitV,
linkcolor=blue, citecolor=blue, urlcolor=blue]{hyperref}

\usepackage[T1]{fontenc}

{%
\setlength{\fboxsep}{0pt}%
\setlength{\fboxrule}{1pt}%
}%
\begin{document}

\title{Unified hydrodynamics for density and spin supersolids}

\author{Ashwath N. Madhusudan}
\affiliation{
	Universit\"{a}t Innsbruck, Fakult\"{a}t f\"{u}r Mathematik, Informatik und Physik, Institut f\"{u}r Experimentalphysik, 6020 Innsbruck, Austria
}

\author{Au-Chen Lee}
\affiliation{Dodd-Walls Centre for Photonic and Quantum Technologies, Dunedin 9054, New Zealand}
\affiliation{Department of Physics, University of Otago, Dunedin 9016, New Zealand}

\author{P. B. Blakie}
\affiliation{Dodd-Walls Centre for Photonic and Quantum Technologies, Dunedin 9054, New Zealand}
\affiliation{Department of Physics, University of Otago, Dunedin 9016, New Zealand}

\author{R. N. Bisset}
\affiliation{
	Universit\"{a}t Innsbruck, Fakult\"{a}t f\"{u}r Mathematik, Informatik und Physik, Institut f\"{u}r Experimentalphysik, 6020 Innsbruck, Austria
}

\begin{abstract}

Two-component supersolids can exhibit in-phase or out-of-phase component density modulations, corresponding to density and spin supersolids, respectively.
We formulate a general hydrodynamic theory for the collective dynamics of multicomponent supersolids, guided by their broken symmetries and conservation laws.
We benchmark it against microscopic calculations for a binary dipolar condensate confined to an infinite tube, finding quantitative agreement for the sound speeds of all three Goldstone branches throughout the density- and spin-supersolid regions.
The resulting relations link collective-mode measurements to elastic coefficients and superfluid fractions.
	
\end{abstract}

\date{\today}
\maketitle

The low-energy excitations of a many-body system provide a direct window into its emergent macroscopic order. In phases with spontaneously broken continuous symmetries, these excitations include Nambu-Goldstone modes \cite{watanabe2012unified,watanabe2020counting}, whose number, dispersion, and symmetry structure reflect the underlying conservation laws and pattern of symmetry breaking. This connection forms the basis of hydrodynamic and effective-field-theory descriptions across a wide range of physical systems, where the relevant degrees of freedom are the slowly varying collective fields associated with conserved quantities and broken symmetries \cite{dubovsky2012effective,brauner2022snowmass}.

A supersolid is a phase of matter in which superfluidity coexists with spontaneously generated crystalline density order. This unique combination establishes a novel many-body setting to explore broken-symmetry physics and collective dynamics. Theoretical work by Andreev and Lifshitz in 1969 \cite{andreev1969quantum} predicted supersolidity and made a general assessment of its hydrodynamic properties, with the hydrodynamic description subsequently being advanced and generalized \cite{saslow1977microscopic,liu1978two,son2005effective,josserand2007coexistence,josserand2007patterns,yoo2010hydrodynamic,heinonen2019quantum,hofmann2021hydrodynamics}.
Recent experiments with ultracold dipolar quantum gases have transformed these longstanding ideas into experimentally accessible many-body systems, exhibiting density-modulated superfluid states \cite{tanzi2019observation,bottcher2019transient,chomaz2019long}.
The associated microscopic theoretical descriptions permit quantitative modeling of experiments and comparison to the established theory of supersolid hydrodynamics
\cite{vsindik2024sound,platt2024sound,rakic2024elastic,poli2024excitations,blakie2025dirac,zawislak2025anomalous,platt2025supersolid,cook2026excitations,poli2026sound,mukherjee2023classical}.

\begin{figure}
	\centering\includegraphics[width=0.48\textwidth]{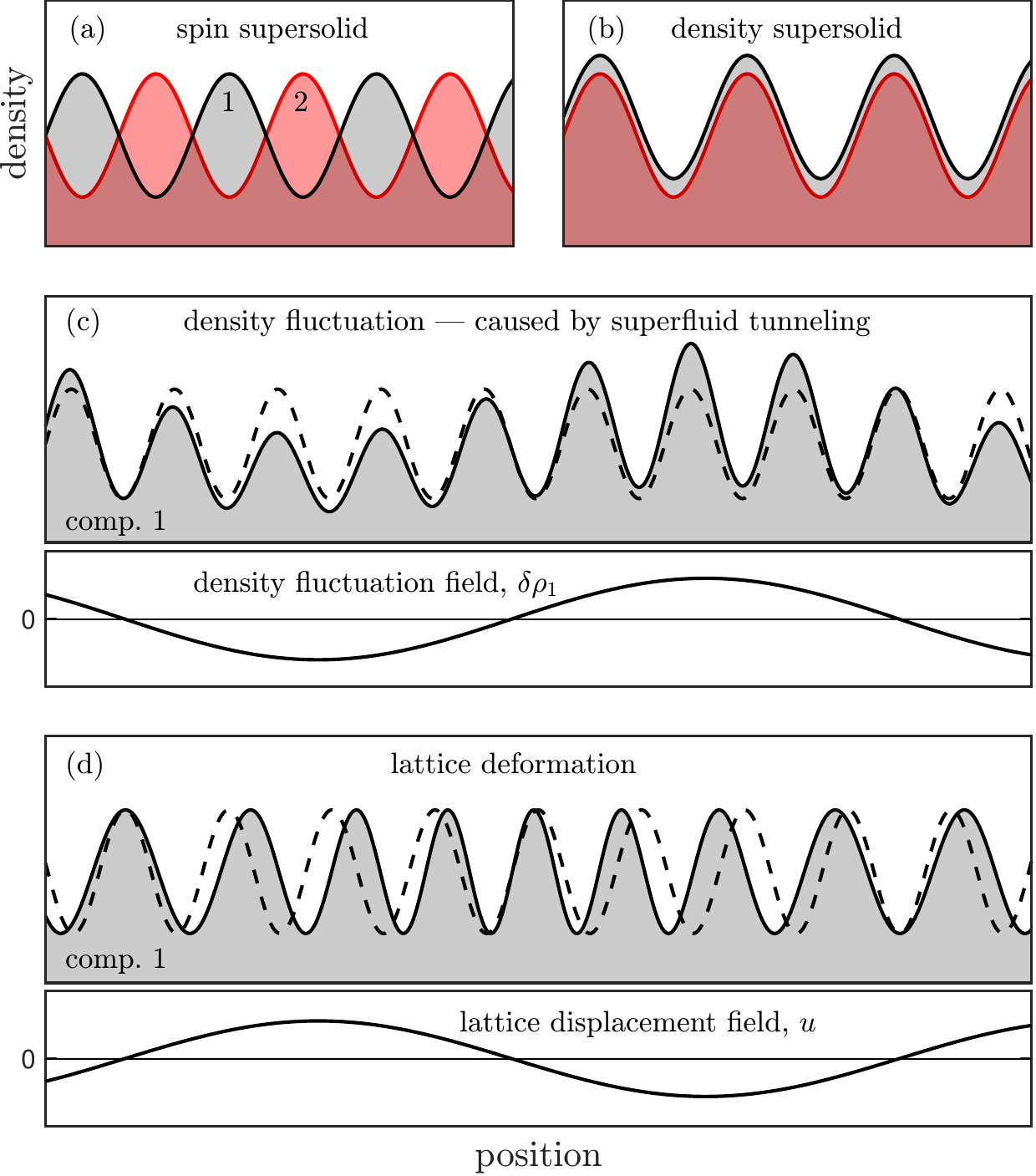}
	\caption{Spin- and density-supersolid schematics and hydrodynamic fields.
		(a) Spin-supersolid; component 1 (black) and component 2 (red) are modulated out of phase.
		(b) Density-supersolid; components are modulated in phase.
		(c) Density fluctuation (upper) and the corresponding slowly varying hydrodynamic field (lower); dashed curves denote ground-state density.
		(d) Lattice deformation (upper) and the associated displacement field (lower).
		We only plot component 1 in (c) and (d) for clarity.
	}
	\label{Fig:Schematic}
\end{figure}

The realization of supersolid phases in ultracold atomic gases has stimulated interest in new types of supersolidity.
Recent experimental advances have expanded the range of accessible multicomponent quantum gases, with long-range interactions \cite{trautmann2018dipolar,durastante2020feshbach,politi2022interspecies,schafer2023realization,kalia2025creation,lecomte2025production}.
Two-component dipolar Bose--Einstein condensates have been proposed to host supersolids in which translational order develops predominantly in either the total-density ($n_1+n_2$) or spin-density ($n_1-n_2$) channel, yielding density supersolids with overlapping (in-phase) sublattices \cite{scheiermann2023catalyzation,scheiermann2025excitation} and spin supersolids with staggered (out-of-phase) sublattices \cite{bland2022alternating,li2022long,kirkby2023spin,arazo2023self,kirkby2024excitations,lee2024excitations} [Fig.~\ref{Fig:Schematic}(a,b)]. Unlike their single-component counterparts, spin supersolids can form away from mechanical instability and hence avoid the high-density quantum-stabilized regime, offering a route to reduced three-body loss, longer lifetimes and larger lattices \cite{bland2022alternating,li2022long}.
At long wavelengths, the one-dimensionally modulated two-component supersolids considered here exhibit three gapless branches arising from the breaking of one translational and two $U(1)$ phase symmetries, together with a gapped optical branch associated with relative sublattice motion \cite{kirkby2024excitations}.

In this letter we develop a general hydrodynamic theory for multi-component supersolids. Benchmarking our theory with a two-component (binary) dipolar supersolid in an infinite-tube geometry, our hydrodynamic predictions for sound speeds agree quantitively with microscopic theory across the breadth of the phase diagram, spanning from spin to density supersolids. We provide analytic predictions when the components are balanced (same intracomponent parameters). We thereby establish a symmetry-based hydrodynamic description of low-energy modes in multicomponent supersolids.

\emph{Formalism}---Hydrodynamic theories of single-component supersolids describe the low-energy collective dynamics in terms of slowly varying density and phase fluctuations coupled to a lattice-displacement field
\cite{yoo2010hydrodynamic,hofmann2021hydrodynamics,vsindik2024sound,platt2024sound}.
Here, we extend this construction to two-component supersolids.

The slow hydrodynamic fields are $\{ \delta \rho_1,\delta \rho_2 ,\phi_1,\phi_2,u \}$,
where $\delta \rho_i$ denote the density fluctuations of component $i=1,\,2$ [Fig.~\ref{Fig:Schematic}(c)],
$\phi_i$ the corresponding phase fluctuations, and $u$ the lattice displacement [Fig.~\ref{Fig:Schematic}(d)].
Intercomponent interactions lock the two sublattice displacements at low energy, so a single displacement field $u$ suffices for both spin and density supersolids.
We introduce the quadratic effective Lagrangian, \footnote{Although we restrict attention here to two components, the symmetry-based construction suggests a natural extension to mixtures with $N$ independently conserved components.}
\begin{align}
	&L = \sum_{i=1}^2 \left\{\frac{m_i \rho^{\rm{n}}_{i}}{2} (v^\mathrm{n} - v_i^\mathrm{sf})^2 - \frac{m_i \rho_i (v_i^{\mathrm{sf}})^{2}}{2} 
	-\hbar \delta \rho_i \frac{\partial \phi_i}{\partial t} \notag \right\} \\
	-& \sum_{i=1}^2\alpha_{u\rho_i} \delta \rho_i \frac{\partial u}{\partial x} 
	- \hspace{-1mm}\sum_{i,j=1}^2 \frac{\alpha_{\rho_i \rho_j}}{2} \delta \rho_i \delta \rho_j
	- \frac{\alpha_{uu}}{2} \hspace{-1mm} \left(\frac{\partial u}{\partial x} \right)^2,
	\label{Eq:2CompLagr}
\end{align}
where $m_i$ are the component masses.
For concreteness, we consider a one-dimensional (1D) crystal (with modulation along $x$) in a three-dimensional, two-component Bose gas confined in an infinitely long harmonic tube \cite{roccuzzo2019supersolid,blakie2020supersolidity};  the generalization to higher-dimensional crystals is straightforward \cite{poli2024excitations}.
The average integrated 1D (line) densities decompose as $\rho_i=\rho^{\rm{sf}}_{i}+\rho^{\rm{n}}_{i}$. 
The superfluid velocity of each component is the gradient of the phase field, $v_i^{\mathrm{sf}} = (\hbar / m_i) \partial_x \phi_i $, while the normal velocity describes lattice motion $v^\mathrm{n} = \partial_t u$.
The elastic coefficients are defined by $(i,j = 1,2)$:
\begin{equation}
	\alpha_{\rho_i \rho_j} = \frac{\partial^2 \mathcal{E}}{\partial \rho_i \partial \rho_j}, \hspace{2mm}
	\alpha_{u\rho_i} = a\frac{\partial^2 \mathcal{E}}{\partial \rho_i \partial a}, \hspace{2mm}
	\alpha_{uu} = a^2 \frac{\partial^2 \mathcal{E}}{\partial a^2},
\end{equation}
where $\mathcal{E}$ is the ground state energy per unit length and $a$ is the lattice constant (End Matter).

We construct the Euler-Lagrange equations from Eq.~(\ref{Eq:2CompLagr}) and seek normal-mode solutions $\delta\rho_i,\phi_i,u \propto e^{i(qx-\omega t)}$ with quasimomentum $q$ and frequency $\omega$.
In the long-wavelength limit the dispersions are linear, $\omega = c q$, yielding three sound modes. 
For generic parameters we obtain the sound speeds by solving the resulting coupled eigenvalue problem numerically. See End Matter for further details, as well as the sound speed equations for the uniform limit simplification.

In the balanced limit ($m_1=m_2\equiv m$, $\rho_1=\rho_2$, and symmetric interactions),
binary supersolid excitations decouple into two independent sectors: density (in-phase fluctuations) and spin (out-of-phase fluctuations) \cite{kirkby2024excitations,lee2024excitations,scheiermann2025excitation}.
We transform to the density/spin basis:
$\rho_{\mathrm{d/s}} = \rho_1\pm\rho_2$, $\delta\rho_{\mathrm{d/s}} = \delta \rho_1 \pm \delta \rho_2$, $\phi_{\mathrm{d/s}} = (\phi_1 \pm \phi_2) /2$,
and, noting that in this limit $\rho_{\rm s}=0$, define the normal and superfluid densities:
$\rho^\mathrm{n}_{\mathrm{d}} = \rho_1^\mathrm{n} + \rho_2^\mathrm{n},
\hspace{1.5mm}
\rho^\mathrm{sf}_{\mathrm{d}} = \rho_1^\mathrm{sf} + \rho_2^\mathrm{sf}$. 
In this basis the nonzero elastic coefficients involving density and spin derivatives are:
\begin{align}
	\alpha_{\rho_{\rm{d}} \rho_{\rm{d}}} = \frac{\partial^2 \xi}{\partial \rho_{\rm{d}}^2}, \quad \alpha_{\rho_{\rm{s}} \rho_{\rm{s}}} = \frac{\partial^2 \xi}{\partial \rho_{\rm{s}}^2}, \quad\alpha_{u\rho_{\rm{d}}} = a\frac{\partial^2 \xi}{\partial \rho_{\rm{d}} \partial a} .
\end{align}  
The spin sector yields a single mode with sound speed
\begin{align}
	& mc_{\rm{s}}^2 = \alpha_{\rho_{\rm{s}} \rho_{\rm{s}}}\rho_{\rm d}^{\rm{sf}}. \label{Eq:SpinSOS}
\end{align}
The density sector hosts two modes with sound speeds
\begin{align}
	&mc_{d\pm}^2 = \frac{1}{2}(a_\Delta \pm \sqrt{a_\Delta^2 - 4b_\Delta}), \notag \\
	&a_\Delta = \rho_{\rm{d}} \alpha_{\rho_{\rm d} \rho_{\rm{d}}}-2\alpha_{u\rho_{\rm{d}}} + \frac{\alpha_{uu}}{\rho_{\rm{d}}^\mathrm{n}}, \notag \\
	&b_\Delta = \frac{\rho_{\rm d}^{\rm{sf}}}{\rho_{\rm d}^{\rm{n}}}(\alpha_{\rho_{\rm d} \rho_{\rm d}}\alpha_{uu} - \alpha_{u\rho_{\rm d}}^2) . \label{Eq:DensSOS}
\end{align}
See End Matter for further details on the derivation.

\emph{System}---We consider a binary dipolar Bose-Einstein condensate confined to an infinite tube trap with longitudinal axis along $x$, and take equal component masses $m_1=m_2\equiv 164u$ (dysprosium) and equal average line densities $\rho_1 = \rho_2 \equiv 750\, \mu m^{-1}$.
The transverse trap frequencies are $(\omega_y,\omega_z) = 2 \pi\times (100,300)$Hz, with the dipoles polarized along $y$, a configuration that favors a linear crystal over more complex radial structures \cite{kirkby2024excitations}.
Beyond-mean-field quantum fluctuations are included via the Lee-Huang-Yang (LHY) correction \cite{bisset2021quantum,smith2021quantum}, which is necessary to stabilize the density-supersolid phase.

To quantitatively benchmark the hydrodynamic theory, we obtain ground states from the extended Gross-Pitaevskii equation and collective excitations, including the sound speeds, from a Bogoliubov-de Gennes (BdG) analysis. Further numerical details are provided in the End Matter.

\begin{figure}[h!]
	\centering
	\includegraphics[width=0.48\textwidth]{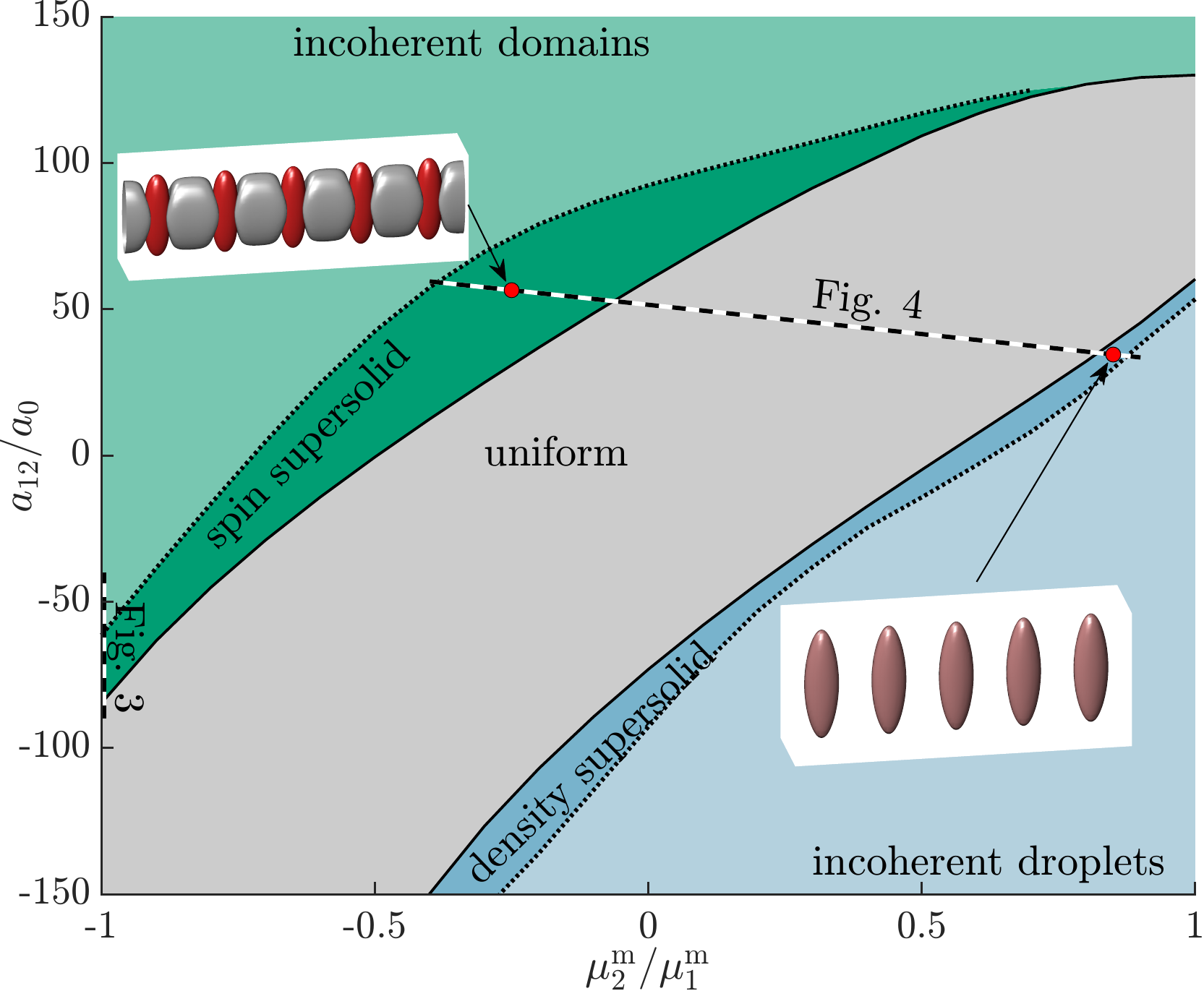}
	\caption{
		Phase diagram showing spin- (green) and density-supersolid (blue) ground state regions as a function of relative dipole moment (horizontal axis) and interspecies scattering length (vertical axis).
		Solid lines denote phase transitions; dotted lines mark where the component-weighted modulation contrast is $\mathcal{C} = 0.95$ (see main text).
		Inset: isosurfaces at 30 \% of peak densities for component 1 (red) and component 2 (gray).
		Striped lines indicate the parameter cuts analyzed in Figs.~\ref{Fig:SpeedsOfSoundSpinSS} and \ref{Fig:SpeedsofSoundimbalanced}.
		Parameters: $a_{11} = a_{22} \equiv 130 a_0, \mu_1^{\rm{m}} \equiv 9.93 \mu_B$, with $\mu_2^{\rm{m}}$ varied.
	}
	\label{Fig:PhaseDiag}
\end{figure}

\emph{Binary supersolid phase diagram}---Figure \ref{Fig:PhaseDiag} shows spin-supersolid (green) and density-supersolid (blue) ground state regions separated by a uniform miscible phase over a broad range of interactions.
The vertical axis is the interspecies s-wave scattering length ($a_{12}$), with the intraspecies scattering lengths ($a_{ii}$) held fixed.
The horizontal axis (dipole imbalance) is tuned by varying the magnetic dipole moment of component 2, $\mu^{\rm{m}}_2$, while keeping $\mu^{\rm{m}}_1$ fixed.

The density-supersolid--uniform boundary is set by softening of the density roton and requires beyond-mean-field stabilization by quantum fluctuations \cite{scheiermann2023catalyzation}.
By contrast, the spin-supersolid--uniform boundary coincides with the miscible--immiscible transition; the unstable spin roton is stabilized by partial demixing within the spin-supersolid phase \cite{bland2022alternating,li2022long}.
A finite dipole imbalance is required to realize the spin supersolid; for our parameters this occurs for $\mu^{\rm{m}}_2/\mu^{\rm{m}}_1 \lesssim 0.7$.

We quantify crystalline order via the per-component contrast $\mathcal{C}_i = (n^{\rm{max}}_i - n^{\rm{min}}_i)/(n^{\rm{max}}_i + n^{\rm{min}}_i)$, where the line densities $n_i(x) = \int dy dz \, |\psi_i|^2$ exhibit periodic minima and maxima.
The component-weighted contrast is $\mathcal{C} = (\rho_1 \mathcal{C}_1 + \rho_2 \mathcal{C}_2)/(\rho_1 + \rho_2)$.
Dotted contours in Fig.~\ref{Fig:PhaseDiag} mark $\mathcal{C}>0.95$, deliniating deeply modulated regimes in which the density at the modulation minima nearly vanishes
and the system crosses over toward the incoherent-domain and incoherent-droplet regimes.
Notably, across our parameter sweep the spin-supersolid region is several times broader than the density-supersolid region, underscoring its greater robustness to parameter variations and making it an ideal target for near-term experiments.

\begin{figure}[h!]
	\centering
	\includegraphics[width=0.48\textwidth]{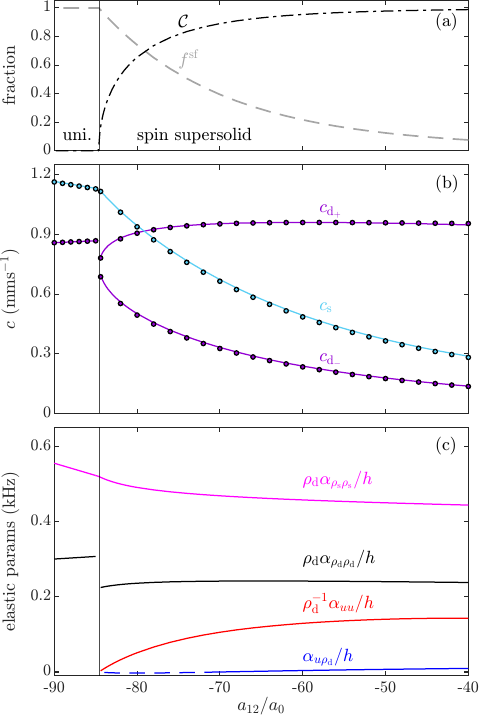}
	\caption{Balanced spin-supersolid: superfluidity, sound and elasticity.
		Hydrodynamic predictions for a balanced mixture along the spin-supersolid cut ($\mu_2^{\rm{m}}/\mu_1^{\rm{m}} = -1$).
		(a) Component-weighted contrast (dot-dashed) and superfluid fraction (dashed).
		(b) Sound speeds from hydrodynamics (solid) and BdG (markers), agreeing to $\lesssim 1\%$. Density (spin) modes are shown as purple (cyan), with a spin--density crossing occuring near $a_{12}/a_0\approx -80$ without hybridization.
		(c) Elastic coefficients, with negative values dashed.
		Parameters as in Fig.~\ref{Fig:PhaseDiag}.
	}
	\label{Fig:SpeedsOfSoundSpinSS}
\end{figure}

\emph{Balanced-mixture hydrodynamics}---Our semianalytic hydrodynamic predictions for balanced mixtures [Eqs.~(\ref{Eq:SpinSOS},\ref{Eq:DensSOS})] can be benchmarked on both supersolid phases of Fig.~\ref{Fig:PhaseDiag} by following vertical cuts at fixed dipole-moment ratio $\mu^{\rm{m}}_2 / \mu^{\rm{m}}_1 =  \pm 1$.
Here we present the spin-supersolid cut ($\mu^{\rm{m}}_2 / \mu^{\rm{m}}_1 = -1$); the density-supersolid phase shows similarly close agreement with BdG and is omitted for brevity.

Figure \ref{Fig:SpeedsOfSoundSpinSS}(a) shows the component-weighted contrast $\mathcal{C}$ and the component-weighted superfluid fraction $f^{\rm sf} = (\rho_1f_1^{\rm sf}+\rho_2f_2^{\rm sf})/(\rho_1+\rho_2)$ versus $a_{12}$ along this cut.
The per-component superfluid fraction is estimated using Leggett's upper bound, $f_i^{\rm sf} = (L/\rho_i) (\int dx/n_i(x) )^{-1}$ \cite{leggett1970can,leggett1998superfluid}.
This trajectory runs from the uniform miscible phase into a broad spin-supersolid window spanning at least $30a_0$.

Figure \ref{Fig:SpeedsOfSoundSpinSS}(b) compares hydrodynamic (solid lines) and BdG (markers) sound speeds in the uniform and spin-supersolid phases, showing remarkable agreement at the level of $\sim$\,1\,\% throughout the parameter cut.
In the fully balanced limit the density and spin sectors decouple, so modes are either spin or density; the exact level crossing near $a_{12}/a_0\approx -80$ reflects the absence of hybridization.
A spin phase mode is present on both sides of the transition and can be tracked continuously across the uniform--supersolid boundary, with a continuous sound speed.
In contrast, the density sector is reorganized at the transition: the density phonon of the uniform state does not carry through as a single branch; in the supersolid it yields a lattice phonon associated with broken translational symmetry (upper) together with a density phase mode (lower), and their velocities are not continuous across the boundary.
Deep in the strongly modulated regime ($a_{12}/a_0 \gtrsim -60$) the upper density branch is clearly the lattice phonon, while both phase modes soften as the superfluid fraction decreases \cite{kirkby2024excitations,lee2024excitations}.

The elastic coefficients are shown in Fig.~\ref{Fig:SpeedsOfSoundSpinSS}(c).
As with the sound speeds, the spin coefficient $\alpha_{\rho_s \rho_s}$ varies continuously across the transition, whereas coefficients involving density derivatives ($\partial/\partial \rho_{\rm d}$) are discontinuous.
The lattice elastic coefficients vanish as the transition is approached from the supersolid side, concomitant with the contrast tending to zero.
Similar to the one-component supersolid \cite{platt2024sound,poli2024excitations}, the density--strain coefficient $\alpha_{u \rho_{\rm d}}$ is small throughout the parameter cut; in some treatments it is neglected altogether \cite{hofmann2021hydrodynamics,zawislak2025anomalous}.

\begin{figure}[h!]
	\centering
	\includegraphics[width=0.48\textwidth]{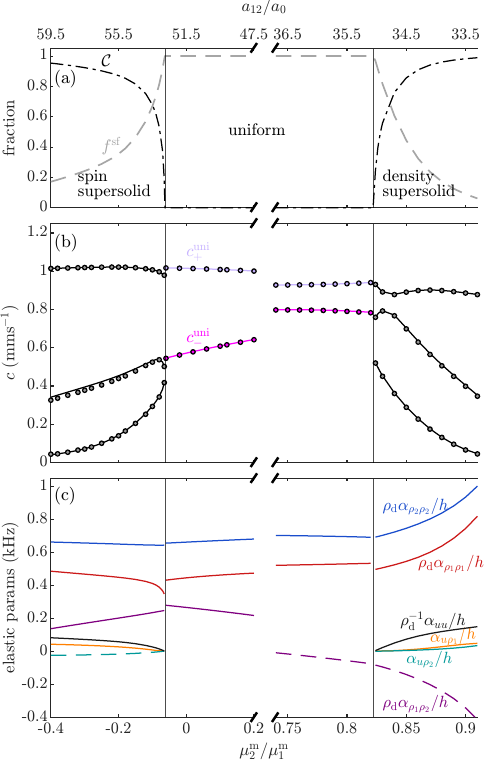}
	\caption{Component-imbalanced supersolids: superfluidity, sound and elasticity.
		Hydrodynamic predictions for component-imbalanced mixtures along a parameter cut that traverses the phase diagram from the spin-supersolid to the density-supersolid.
		(a) Component-weighted contrast and superfluid fraction.
		(b) Sound speeds from hydrodynamics (solid) and BdG (markers).
		(c) Elastic coefficients, with negative values dashed.
		Parameters as in Fig.~\ref{Fig:PhaseDiag}.	
	}
	\label{Fig:SpeedsofSoundimbalanced}
\end{figure}

\emph{General hydrodynamics across the phase diagram}---Figure \ref{Fig:SpeedsofSoundimbalanced} follows a diagonal parameter cut for imbalanced components from the spin-supersolid, through the uniform phase, and into the density-supersolid (see Fig.~\ref{Fig:PhaseDiag}); it uses the same panel layout as Fig.~\ref{Fig:SpeedsOfSoundSpinSS}.
We numerically compute the hydrodynamic sound speeds as outlined in End Matter.
Figure \ref{Fig:SpeedsofSoundimbalanced}(b) shows these predictions as solid lines alongside BdG results (markers), with excellent quantitative agreement across all three phases.
Because the component imbalance couples spin and density channels, no exact level crossings are observed, and all sound speeds are discontinuous at both transitions.

Figure \ref{Fig:SpeedsofSoundimbalanced}(c) shows the elastic coefficients: the lattice modulus ($\alpha_{uu}$) and density--lattice couplings ($\alpha_{u\rho_i}$) soften to zero at the phase boundaries---vanishing in the uniform segment---as in the balanced case [Fig.~\ref{Fig:SpeedsOfSoundSpinSS}(c)].
The density--density coefficients ($\alpha_{\rho_i\rho_j}$) remain finite across the cut but are discontinuous at the transitions.	
The off-diagonal $\alpha_{\rho_1\rho_2}$ changes sign---positive in the spin-supersolid and part of the uniform phase, negative thereafter---consistent with increasingly attractive inter-component dipolar interactions as $\mu_2^{\rm m}$ increases (with $\mu_1^{\rm m}$ held fixed).
Compared with the balanced case, the density--lattice coulings $\alpha_{u\rho_i}$ play a relatively larger role: if we artificially set $\alpha_{u\rho_i}=0$, the hydrodynamic sound speed predictions in Fig.~\ref{Fig:SpeedsofSoundimbalanced}(b) are modified by $\sim\,$10\,\% at both $\mu_2^{\rm m}/\mu_1^{\rm m} \approx -0.4$ and $0.9$ (not shown).

\emph{Conclusions}---We have developed a hydrodynamic framework for multicomponent supersolids that connects densities, superfluid fractions, and elastic degrees of freedom to the long-wavelength excitations.
While this description is independent of the microscopic interaction details---relying only on symmetry and conservation laws---we have demonstrated its utility with a binary dipolar mixture and find quantitative agreement for all sound speeds between our hydrodynamic predictions and microscopic Bogoliubov--de Gennes calculations throughout the supersolid phase diagram.
For balanced mixtures, we have derived closed-form analytic expressions for the sound speeds in terms of the elastic coefficients, whereas for imbalanced mixtures we compute these relations numerically.

Our work provides a route to experimentally access elastic coefficients and superfluid fractions through measurements of long-wavelength mode speeds using established techniques \cite{guo2019low,tanzi2019supersolid,petter2021bragg,biagioni2024measurement,vsindik2024sound}.
Our mapping of the phase diagram of a binary dipolar mixture reveals both spin- and density-supersolid regions separated by an unmodulated uniform phase.
The spin-supersolid parameter regime is particularly broad and thus robust to parameter variations.
Moreover, because the spin-supersolid does not rely on quantum-fluctuation stabilization, it can occur at lower densities.
This is expected to reduce inelastic losses and enable larger, longer-lived supersolid lattices \cite{bland2022alternating,li2022long}. These features make it an attractive target for future experiments in binary dipolar mixtures \cite{trautmann2018dipolar,durastante2020feshbach,politi2022interspecies,schafer2023realization,kalia2025creation,lecomte2025production}.

Because the construction of the hydrodynamic theory relies only on symmetries, it applies directly to other platforms with long-range-interactions, including bilayer dipolar molecule gases \cite{baranov2012condensed}, multicomponent condensates in multimode cavities \cite{mivehvar2021cavity}, and Rydberg-dressed supersolids with soft-core interactions \cite{pomeau1994dynamics,henkel2010three,cinti2010supersolid}.
Extending the framework to 2D/3D supersolid lattices would capture anisotropic longitudinal and transverse phonons and their coupling to superfluid flow, and a finite-temperature generalization would include a thermal normal component, entropy/heat transport, and dissipation.

{\it Acknowledgements}---We acknowledge stimulating discussions with Andrew Underwood, Rene R\"ohrs, Alessio Recati, Sandro Stringari, Elena Poli, Natalia Masalaeva and Hannah Geiger.
This research was funded in whole or in part by the Austrian Science Fund (FWF) [10.55776/P36850].
For open access purposes, the author has applied a CC BY public copyright licence to any author accepted manuscript version arising from this submission.


%


\appendix

\onecolumngrid     
\newpage
\section*{End Matter}

\twocolumngrid

\setcounter{equation}{0}
\renewcommand{\theequation}{A\arabic{equation}}

\newcommand{\smallsum}{\mathop{\textstyle\sum}}
\emph{Hydrodynamic theory}---Using the Lagrangian [Eq.~(\ref{Eq:2CompLagr})], the Euler-Lagrange (EL) equations are,
\begin{align}
	\frac{\partial L}{\partial (\delta \rho_i)} - \frac{\partial }{\partial x} \left(\frac{\partial L}{\partial (\partial_x \delta \rho_i)}\right) - \frac{\partial }{\partial t} \left(\frac{\partial L}{\partial (\partial_t \delta \rho_i)}\right) &= 0, \notag \\   
	\frac{\partial L}{\partial (\phi_i)} - \frac{\partial }{\partial x} \left(\frac{\partial L}{\partial (\partial_x \phi_i)}\right) - \frac{\partial }{\partial t} \left(\frac{\partial L}{\partial (\partial_t \phi_i)}\right) &= 0, \notag \\
	\frac{\partial L}{\partial (u)} - \frac{\partial }{\partial x} \left(\frac{\partial L}{\partial (\partial_x u)}\right) - \frac{\partial }{\partial t} \left(\frac{\partial L}{\partial (\partial_t u)}\right) &= 0,    
\end{align}
where $i=1,2$. Simplifying the EL equations gives
\begin{align}
&\hbar \partial_t \phi_1
+ \alpha_{\rho_1 u} \partial_x u
+ \alpha_{\rho_1 \rho_1} \delta\rho_1
+ \alpha_{\rho_1 \rho_2} \delta\rho_2
= 0 \label{Eq:EL} \\
&\hbar \partial_t \phi_2
+ \alpha_{\rho_2 u} \partial_x u
+ \alpha_{\rho_2 \rho_2} \delta\rho_2
+ \alpha_{\rho_1 \rho_2} \delta\rho_1
= 0 \notag \\
&\rho_1 \tfrac{\hbar^2}{m_1} \partial_x^2 \phi_1
+ \hbar \rho_1^{\rm n} \left( \partial_{tx} u
 - \tfrac{\hbar}{m_1} \partial_x^2 \phi_1 \right)
+ \hbar \partial_t \delta\rho_1
= 0 \notag \\
&\rho_2 \tfrac{\hbar^2}{m_2} \partial_x^2 \phi_2
+ \hbar \rho_2^{\rm n} \left( \partial_{tx} u
 - \tfrac{\hbar}{m_2} \partial_x^2 \phi_2 \right)
+ \hbar \partial_t \delta\rho_2
= 0 \notag \\
&\smallsum_{i=1}^2 \hspace{-0.5mm} \alpha_{u \rho_i} \partial_x \delta\rho_i
+ \alpha_{uu} \partial_x^2 u
- \smallsum_{i=1}^2 \hspace{-0.5mm} m_i \rho_i^{\rm n} \hspace{-0.5mm} \left( \partial_t^2 u
 - \tfrac{\hbar}{m_i} \partial_{tx} \phi_i \right)
= 0 . \notag
\end{align}
We assume normal mode solutions of the form $\delta\rho_i,\phi_i,u \propto e^{i(qx-\omega t)}$ and then rewrite Eqs.~(\ref{Eq:EL}) in the form $MX = 0$, where
\begin{align}
	X = \begin{pmatrix}
		\delta \rho_1 & \delta \rho_2 & \phi_1 & \phi_2 & u
	\end{pmatrix}^{T} ,
\end{align}
and the dynamical matrix $M$ is	
\begin{align}
M = \left(\begin{smallmatrix}
\alpha_{\rho_1\rho_1} & \alpha_{\rho_1\rho_2} & -i\hbar\omega & 0 & iq\,\alpha_{u\rho_1} \\
\alpha_{\rho_1\rho_2} & \alpha_{\rho_2\rho_2} & 0 & -i\hbar\omega & iq\,\alpha_{u\rho_2} \\
-i\hbar\omega & 0 & -\tfrac{\rho_1^{\mathrm{sf}}\hbar^2 q^2}{m_1} & 0 & \hbar\omega q\,\rho_1^{\mathrm{n}} \\
0 & -i\hbar\omega & 0 & -\tfrac{\rho_2^{\mathrm{sf}}\hbar^2 q^2}{m_2} & \hbar\omega q\,\rho_2^{\mathrm{n}} \\
iq\,\alpha_{u\rho_1} & iq\,\alpha_{u\rho_2} & \hbar\omega q\,\rho_1^{\mathrm{n}} & \hbar\omega q\,\rho_2^{\mathrm{n}} & A\omega^2 - \alpha_{uu}q^2
\end{smallmatrix}\right) ,
\end{align}
where $A = m_1\rho_1^{\rm n} + m_2\rho_2^{\rm n}$.
Non-trivial solutions require the determinant $\Delta_{M} = 0$.
In the general imbalanced case, getting an analytical expression for the speeds of sound is challenging. We convert the problem to a quadratic eigenvalue problem and solve it numerically.

\emph{Uniform-limit sound speeds}---For the imbalanced case in the unmodulated limit the elastic coefficients associated with the lattice displacement are zero and the sound speeds are 
\begin{align}
	c^{\rm{uni}}_\pm &= \sqrt{\frac{1}{2m_1 m_2} \left(a_{\rm{u}} \pm \sqrt{b_{\rm{u}}} \right)}, \label{Eq:UniSounds} \\
	a_{\rm{u}} &= m_2 \alpha_{\rho_1\rho_1} \rho_1  + m_1 \alpha_{\rho_2\rho_2} \rho_2 , \notag \\
	b_{\rm{u}} &= \left(m_2\alpha_{\rho_1\rho_1} \rho_1 - m_1\alpha_{\rho_2\rho_2} \rho_2\right)^2 + 4m_1 m_2\alpha_{\rho_1\rho_2}^2\rho_1\rho_2 .	 \notag
\end{align}

\emph{Balanced-supersolid sound speeds}---In the limit of balanced components, we transform to the density/spin basis:
\begin{align}
\delta\rho_{\mathrm{d/s}} &= \delta \rho_1 \pm \delta \rho_2, \quad \phi_{\mathrm{d/s}} = (\phi_1 \pm \phi_2) /2, \\
\rho_{\mathrm{d/s}} &= \rho_1\pm\rho_2, \quad \rho^\mathrm{n}_{\mathrm{d/s}} = \rho_1^\mathrm{n} \pm \rho_2^\mathrm{n}, \quad \rho^\mathrm{sf}_{\mathrm{d/s}} = \rho_1^\mathrm{sf} \pm \rho_2^\mathrm{sf} , \notag
\end{align}
where $\rho_{\rm s} = \rho_{\rm s}^{\rm n} = \rho_{\rm s}^{\rm sf}=0$.

Spin excitations cause the hydrodynamic fields to fluctuate antisymmetrically between the components:
\begin{equation}
\delta\rho_1 = - \delta\rho_2, \quad \phi_1 = -\phi_2 ,
\end{equation}
with the Lagrangian reducing to	
\begin{align}
	L_{\rm s} = &-\hbar \delta \rho_{\rm{s}} \partial_t \phi_{\rm{s}} - \rho_{\rm d} \tfrac{\hbar^2}{2m}\left(\partial_x \phi_{\rm{s}}\right)^2 - \tfrac{1}{2} \alpha_{\rho_{\rm s}\rho_{\rm s}} \delta \rho_{\rm{s}}^2 \label{Spin Lagrangian}  \\
	& - \tfrac{1}{2}\alpha_{uu}\left( \partial_x u \right)^2 + \tfrac{1}{2}m \rho_{\rm d}^{\rm n}\left(\left(\partial_t u\right)^2 + \left(\tfrac{\hbar}{m}\partial_x \phi_{\rm{s}}\right)^2 \right) . \notag
\end{align}
The dynamical matrix corresponding to $L_{\rm s}$ is
\begin{align}
	M_{\rm s} = 
	\begin{pmatrix}
		\alpha_{\rho_{\rm s}\rho_{\rm s}}&-i\hbar \omega & 0\\
		-i\hbar \omega & -\hbar^2 q^2 \rho_{\rm d}^{\rm{sf}}/m & 0 \\
		0 & 0 & m\rho_{\rm d}^{\rm n}\omega^2 -  \alpha_{uu}q^2
	\end{pmatrix} , \label{Spin Dynamical Matrix}
\end{align}
and
\begin{align}
	X_{\rm s} = 
	\begin{pmatrix}
		\delta \rho_{\rm{s}} & \phi_{\rm{s}} & u
	\end{pmatrix} ^{T} .
\end{align}
The decoupling of $u$ from the other hydrodynamic fields means that spin excitations do not deform the lattice in the balanced limit.
Non-trivial solutions require the determinant $\Delta_{M_{\rm s}} = 0$, so $\Delta_{M_{\rm s}} = \hbar^2(\omega^2 - c_{\rm s}^2q^2)$ yields $c_{\rm s}$ given in Eq.~(\ref{Eq:SpinSOS}).

Density excitations cause the hydrodynamic fields to fluctuate symmetrically between the components:
\begin{equation}
\delta\rho_1 = \delta\rho_2, \quad \phi_1 = \phi_2 ,
\end{equation}
and thus the Lagrangian [Eq.~(\ref{Eq:2CompLagr})] reduces to
\begin{align}
	L_{\rm d} = &-\hbar \delta \rho_{\rm d} \partial_t \phi_{\rm d} - \tfrac{\alpha_{\rho_{\rm d} \rho_{\rm d}}}{2} \delta\rho_{\rm d}^2  - \tfrac{\alpha_{uu}}{2}\left(\partial_x u\right)^2 \nonumber \\
	&- \alpha_{u\rho_{\rm d}}\delta \rho_{\rm d} \left(\partial_x u\right) - \rho_{\rm d} \tfrac{\hbar^2}{2m}\left(\partial_x \phi_{\rm d}\right)^2 \nonumber \\
	&+ \tfrac{1}{2}m\rho_{\rm d}^{\rm n} \left(\partial_tu - \tfrac{\hbar}{m}\partial_x \phi_{\rm d} \right)^2 ,
	\label{Density Lagrangian}
\end{align}
giving the EL equations:
\begin{align}
	&M_{\rm d} = 
	\begin{pmatrix}
		\alpha_{\rho_{\rm d}\rho_{\rm d}}&-i\hbar \omega& iq\alpha_{u\rho_{\rm d}} \\
		i\hbar \omega & \hbar^2 q^2 \rho_{\rm d}^{\rm sf}/m & -\hbar \omega q \rho_{\rm d}^{\rm n} \\
		iq\alpha_{u\rho_{\rm d}} & \hbar \omega q \rho_{\rm d}^{\rm n} & m\rho_{\rm d}^{\rm n} \omega^2 - \alpha_{uu}q^2
	\end{pmatrix}, \label{Density Dynamical Matrix} \\
	&X_{\rm d} = 
	\begin{pmatrix}
		\delta \rho_{\rm d} & \phi_{\rm d} & u
	\end{pmatrix}^{T} ,
\end{align}
with determinent
\begin{align}
	\Delta_{M_{\rm d}} = m \hbar^2 \rho_{\rm d}^{\rm n}(\omega^2 - c_{{\rm d}_-}^2q^2)(\omega^2 - c_{{\rm d}_+}^2q^2) ,
\end{align}
where $c_{{\rm d}\pm}$ are given in Eq.~(\ref{Eq:DensSOS}).

\emph{Gross--Pitaevskii formalism}---The energy functional can be written for a three-dimensional system as
\begin{align}
	&E = \hspace{-0.5mm} \int \hspace{-1mm} d \textbf{x} \sum_{i = 1}^{2} \psi^*_i(\textbf{x},t) \hspace{-0.5mm} \left( \frac{-\hbar^2 \nabla^2}{2m_i} + V_i(\textbf{x}) \right) \hspace{-0.5mm} \psi_i(\textbf{x},t) \label{EQ:EFun}  \\ 
	& + \frac{1}{2}\int \hspace{-1mm} d \textbf{x} \int \hspace{-1mm} d \textbf{x} ^\prime \sum_{i,j = 1}^{2} n_i(\textbf{x}) V_{ij}(\textbf{x} - \textbf{x}^{\prime}) n_j(\textbf{x}^\prime) + \int \hspace{-1mm} d \textbf{x} \, \varepsilon_{\textrm{QF}} , \notag
\end{align}
where $V_i(\textbf{x}) = m_i(\omega_y^2 y^2 + \omega_z^2 z^2)/2$ is the trapping term, $\psi_i(\textbf{x},t)$ is the wavefunction of component $i$, and $n_i(\textbf{x}) = |\psi_i(\textbf{x},t)|^2$ is its density. The interaction term is
\begin{align}
	V_{ij}(\textbf{x} - \textbf{x}^{\prime}) = 	g_{ij} \delta(|\textbf{x} - \textbf{x}^{\prime}|) + \frac{3g_{ij}^{\rm{dd}}}{4 \pi} \left( \frac{1-3 \rm{cos}^2 \theta}{|\textbf{x} - \textbf{x}^{\prime}|^{3}} \right),
\end{align}
with contact coupling $g_{ij} = 4\pi a_{ij}\hbar^2/m$, and dipolar coupling $g_{ij}^{\rm{dd}} = 4 \pi a_{ij}^{\rm{dd}} \hbar^2/m = \mu_0 \mu^{\rm{m}}_i \mu^{\rm{m}}_j/3$, where $a_{ij}$ are the s-wave scattering lengths, $a_{ij}^{\rm{dd}}$ the dipole lengths, $\mu^{\rm{m}}_i$ the dipole moments of the components, and $\theta$ the angle between the vector $\textbf{x} - \textbf{x}^{\prime}$ and the polarization direction.

We include the quantum-fluctuation energy density $\varepsilon_{\textrm{QF}}$ as formulated in Refs.~\cite{bisset2021quantum,smith2021quantum} and evaluate it under the symmetric-mixture approximation (applied within $\varepsilon_{QF}$ only; see also \cite{smith2021approximate}):
$n_1(\mathbf{x}), n_2(\mathbf{x}) \to \bar{n} \equiv   (n_1(\mathbf{x}) + n_2(\mathbf{x}))/2$;
$U_{11}, U_{22} \to \overline{U} \equiv \sqrt{U_{11} U_{22}}$,
with $U_{ij}(\mathbf{k}) = g_{ij}^{\rm{s}} + g_{ij}^{\rm{dd}}(3 \cos^2 \theta_k - 1)$ the Fourier transform of the total interaction potential, where $\theta_k$ is the angle between $\mathbf{k}$ and the dipole polarization axis. The interspecies coupling $U_{12}(\mathbf{k})$ is left unchanged.
This approximation is excellent for nearly balanced mixtures (including the density-supersolid regime considered here). We also retain $\varepsilon_{\rm QF}$ in the spin-supersolid regime; at the lower densities there its contribution is negligible, but keeping it simplifies comparisons across regimes and underscores the model independence of our hydrodynamic theory.
We then obtain
\begin{align}
	&\varepsilon_{\rm{QF}} = \frac{2}{5}K \left(n_1(\mathbf{x}) + n_2(\mathbf{x}) \right)^{5/2} \\
	&K = \frac{m^{3/2}}{3 \sqrt{2} \pi^2 \hbar^3} \int_0^{\pi/2} \hspace{-1mm} d\theta_k \sin \theta_k\sum_\pm \left(\overline{U} \pm U_{12}  \right)^{5/2} . \notag
\end{align}

The Gross-Pitaevskii equations derived from (\ref{EQ:EFun}) are
\begin{align}
	&i \hbar \frac{\partial \psi_i(\textbf{x},t)}{\partial t} = \mathcal{L}_i \psi_i(\textbf{x},t) ,
\end{align}
where the GPE operator is given by
\begin{align}
	\mathcal{L}_i = \frac{-\hbar^2 \nabla^2}{2m_i} + V_i(\textbf{x}) + \sum_{j=1}^2 \int d\textbf{x}^\prime V_{ij}(\textbf{x} - \textbf{x}^{\prime})n_j(\textbf{x}^{\prime}) + \Delta \mu, 
\end{align}
with \begin{align}
	\Delta \mu  = K\left(n_1(\textbf{x}) + n_2(\textbf{x}) \right)^{3/2}. 
\end{align}

\emph{Bogoliubov--de Gennes formalism}---To obtain the excitations of GPE ground states, we use the BdG ansatz
\begin{align}
	\Psi_i(\textbf{x},t) = \left\lbrace \psi_i^0(\textbf{x}) + \lambda\left[u_i(\textbf{x})e^{-i \omega t} - v_i^*(\textbf{x})e^{i \omega^* t}\right] \right\rbrace e^{-i \mu_i t/ \hbar}\label{BdG ansatz}
\end{align}
where $\psi_i^0(\textbf{x})$ are the ground state solutions, $\mu_i$ are the chemical potentials, and $u_i(\textbf{x}),v_i(\textbf{x})$ are the Bogoliubov amplitudes.
We use this ansatz to linearize the GPE equations and obtain the following eigenvalue equations:
\begin{align}
	&\begin{pmatrix}
		&\mathbf{L} + \mathbf{X} + \mathbf{F} &-\mathbf{X} - \mathbf{F} \\
		&\mathbf{X} + \mathbf{F} & -\mathbf{L} - \mathbf{X} - \mathbf{F}
	\end{pmatrix}
	\begin{pmatrix}
		\mathbf{u}\\
		\mathbf{v}
	\end{pmatrix}
	= \epsilon\begin{pmatrix}
		\mathbf{u}\\
		\mathbf{v}
	\end{pmatrix} , \\
	& \qquad \mathbf{L} = \begin{pmatrix}
		&\mathcal{L}_1 - \mu_1 &0 \\
		&0 & \mathcal{L}_2 - \mu_2 
	\end{pmatrix} , \\
	& \qquad \mathbf{u} = \begin{pmatrix}
		u_1& u_2
	\end{pmatrix}^{T} ,\mathbf{v} = \begin{pmatrix}
		v_1& v_2
	\end{pmatrix}^{T} ,
\end{align}
where, taking $\psi_i^0$ to be real,
\begin{align}
	&(X_{ij}\chi_j)(\textbf{x}) = \psi_i^0(\textbf{x}) \int d \textbf{x}^\prime \psi_j^0(\textbf{x}^\prime) V_{ij}(\textbf{x} - \textbf{x}^\prime) \chi_j(\textbf{x}^\prime),\\ 
	&(F_{ij} \chi_j)(\textbf{x}) = \frac{3K}{2} \sqrt{n_1(\textbf{x}) + n_2(\textbf{x})} \psi_i^0(\textbf{x}) \psi_j^0(\textbf{x})\chi_j(\textbf{x}).
\end{align}

These equations can effectively be solved by writing the Bogoliubov amplitudes in terms of the Bloch amplitudes $\tilde{u}_i(\textbf{x}),\tilde{v}_i(\textbf{x})$, which are periodic in the unit cell,
\begin{align}
	u_i(\textbf{x}) = \tilde{u}_i(\textbf{x}) e^{i q_z z}, \quad
	v_i(\textbf{x}) = \tilde{v}_i(\textbf{x}) e^{i q_z z} ,
\end{align}
where $q_z$ is the quasimomentum.

\end{document}